# A Novel Approach Based Deep RNN Using Hybrid NARX-LSTM Model For Solar Power Forecasting

Mohamed Massaoudi[1,2], Ines Chihi[3], Lilia Sidhom[3], Mohamed Trabelsi[2,4], Shady S. Refaat[2], Fakhreddine S. Oueslati[1]

[1]Unité de Recherche de Physique des Semi-Conducteurs et Capteurs, Carthage University, Tunis, Tunisia

[2]Department of Electrical and Computer Engineering, Texas A&M University at Qatar, Doha, Qatar

[3]LAboratory of Research in Automation (L.A.R.A), National Engineering School of Tunisia, Tunisia

[4]Department of electronics and communications engineering, Kuwait College of Science, Kuwait

{mohamed.massaoudi, shady.khalil}@qatar.tamu.edu;{Ines.Chihi, Lilia.Sidhom}@enib.rnu.tn;
m.trabelsi@kcst.edu.kw; fakhreddine.oueslati@fst.rnu.tn

*Abstract*—The high variability of weather parameters is making photovoltaic energy generation intermittent and narrowly controllable. Threatened by sudden discontinuity between the load and the grid, energy management for smart grid systems highly require an accurate PV power forecasting model. In this regard, Nonlinear autoregressive exogenous (NARX) is one of the few potential models that handle time series analysis for long-horizon prediction. This later is efficient and high-performing. However, this model often suffers from the vanishing gradient problem which limits its performances. Thus, this paper discus NARX algorithm for long-range dependencies. However, despite its capabilities, it has been detected that this model has some issues coming especially from the vanishing gradient. For the aim of covering these weaknesses, this study suggests a hybrid technique combining long short-term memory (LSTM) with NARX networks under the umbrella of Evolution of recurrent systems with optimal linear output (EVOLINO). For the sake of illustration, this new approach is applied to PV power forecasting for one year in Australia. The proposed model enhances accuracy. This made the proposed algorithm outperform various benchmarked models.

## I. Introduction

Clean energy transition toward renewable sources (RS) is becoming a necessity in the twenty-one century. This goal has been cited in France summit agreement. Obviously, RS ensures sustainable development and overcome the nature devastation issues. Within this vein, solar energy from photovoltaic plants is leading this transition. However, it has been noticed that RS has a major drawback in terms of stability and power quality. Photovoltaic panels are continuously disturbed by weather conditions such as clouds, wind speed, and temperature. Also, it must be mentioned that the key element for energy generation which is the irradiation is unavailable during the night hours.

Thus, PV power forecasting models are proposed to estimate the power generated from solar energy and then ensure unit commitment and budget planning. Therefore, time series forecasting (TSF) is becoming a dynamic research area supported by the exponential rise of big data led by the exponential growth of the internet. The latter consequence gives birth to new accurate techniques [1].In this respect, TSF interprets the behavior of some variables that continuously change over time to reconstitute a clear vision about future values [2]. i.e. understanding the past to estimate the future. TSF is frequently used in econometrics, statistical analysis [3], finance [4], weather forecasting [5], [6] and many other uncountable applications. In this context, forecasting the photovoltaic power is mainly predicted through two methods: PV power is predicted via numerical weather prediction (NWP) using mathematical modeling from the natural phenomenon or satellite observations. The data assimilation technique analyses the patterns of satellite information and the actual climate conditions[7][8][9]. The data acquisition from the first type of forecasting uses indirect measures i.e. it indicates how much the weather changes the environment[10]. This leads to the second type of power forecasting which is interpreted in this paper. This type consists of using direct measurement from ecological elements in the process[11]. The prediction analyses weather database to determine the next photovoltaic power. It had been approved that the second approach is more precise[12]. Moreover, it gets more attention due to the development of advanced algorithms to get higher accuracy for a longer time horizon. This leads us to three categories for forecasting horizons. Short term prediction includes an hourly prediction for sudden dispatching. Medium domain forecasting is a daily estimation of the photovoltaic output for maintenance planning. The last type is long term prediction that lasts for years. The usage of the aforementioned type involves budget planning and project investment[13],[14][15].

Among all these classes, forecasting methods were proposed to analyze time-series data. performant models can predict solar power with less computational computing, less features input parameters and for longer time dependencies. Time series models for supervised learning includes statistical approaches, ensemble methods and artificial neural networks (ANN)[16]. A recurrent neural network as one of deep ANN is used to forecast PV power for short period of time. It has been detected that the vanishing gradient limits the time horizon[17]. In the same direction NARX networks suffer from the same problem. To overcome these issues, LSTM cells are proposed for RNN model to capt the important information for a longer time. In this regard, hybrid models were widely used in forecasting and were a source of inspiration for researchers to combine different predictors. This fusion presents a robust model that can outperform individual methods[18]–[20]. From that perspective, this paper proposes a hybrid method form NARX and LSTM-RNN.

Within this framework, the contributions of this paper are resumed in three folds:

1. The first part investigates NARX network architecture.
2. The second part proposes a hybrid method for non-stationary time series prediction composed of LSTM cells and NARX networks. The fusion between these two approaches strengthens the predictor accuracy.
3. The evaluation of the new approach is done through real datasets targeting the PV power forecasting in medium/long-term dependencies.

## II. FORECASTING MODELS OVERVIEW

Forecasting algorithms analyze time-series patterns. These methods conclude various univariate and multivariate time series (UTS) [21]-[22]. From the basic methods such as exponential smoothing, Moving average(MA), autoregressive (AR) to the fusion between them[23]-[24]-[25]. Taken as examples, autoregressive moving average (ARMA) model and autoregressive integrated moving average (ARIMA) analyses a stationary time-series database to extracts statistical information from them[26]. These models are well known in the very short term forecasting due to their ability to extract the output power with an acceptable level of accuracy[27]. Alternatively, the inputs aren't always stationary so this method is less accurate due to the high variability of the weather parameters. AR $(p)$, MA $(q)$, ARMA $(p,q)$ and ARIMA $(p,q)$ equations are written in Eq. (1)-(4).

$$Y_t = \mu + \sum_{k=1}^{p} \Phi_k Y_{t-k} + \epsilon_t \qquad (1)$$

$$Y_t = \mu + \epsilon_t + \sum_{k=1}^{q} \Phi_k \epsilon_{t-k} \qquad (2)$$

$$Y_t = \mu + \epsilon_t + \sum_{k=1}^{q} \Phi_k Y_{t-k} + \epsilon_t + \sum_{k=1}^{q} \theta_k \epsilon_{t-k} \qquad (3)$$

$$Y_t = \mu + \sum_{k=1}^{q} \Phi_k Y_{t-k} + \sum_{k=1}^{q} \theta_k \epsilon_{t-k} \qquad (4)$$

With:

| | |
|---|---|
| $p$ | Autoregressive model order |
| $q$ | Moving average model order |
| $\emptyset$ | Autoregressive parameter |
| $\theta$ | Moving average parameter |
| $\mu$ | Mean value |
| $k$ | Initial value |
| $Y_{t-k}$ | Observed value at time $t-k$ |
| $\epsilon_{t-k}$ | Forecast error at time $t-k$ |

However, these methods are unable to follow nonlinear TS dependencies. Therefore, Nonlinear autoregressive (NAR) and Nonlinear autoregressive with exogenous inputs (NARX) is proposed. The contribution of these latter is their ability to deal efficiently with dynamic features [28], [29].

Moving to machine learning technics presented by the famous artificial neural network (ANN) model which provides accurate results. Given the non-linearity of the meteorological data, ANN is self-adaptive, highly efficient and proven its performance with weather parameters forecasting [30]. Cognitive scientists led by John Hopfield suggest recurrent neural networks (RNN). This model is known as one of the most powerful algorithms for its robustness in learning from past values. Deep learning is also involved in TS prediction used with the big data availability. However, ANN suffers from long computational time in the training phase. Moreover, this method requires a complete database for the training which has a major impact on output accuracy [31].

On the other side, hybrid models are a combination of two or more prediction models. This enhances accuracy since the feature of each model will be transferred. In the literature, Nima Amjady et al. introduced various hybrid methods for load, and power forecasting based on statistical and NN algorithms with the aim of an efficient energy management system [32],[33]. H. Nazaripouya et al. created a TS model of Solar Power Forecasting using Hybrid Wavelet-ARMA-NARX[19], Huaizhi Wang et al. in 2017 proposed a new hybrid method for deterministic PV power forecasting based on wavelet transform (WT) and deep convolutional neural network (DCNN)[34], Yordanos Kassa Semero et al. suggests using a GA-PSO-ANFIS approach in PV power forecasting based feature selection strategy[35] Fang Liu et al. proposed Takagi–Sugeno fuzzy model-based approach In PV power short-range prediction in 2017.[36] and Ji Wu and Chee Keong Chan proposed a novel hybrid model composed of ARMA and TDNN for hourly solar radiation[19].

### I. NARX NEURAL NETWORK

In the literature, a Nonlinear autoregressive network with exogenous inputs (NARX) is a part of discrete-time Nonlinear systems. This hybrid design involves the genetic algorithm (GA)-based optimization technique in the optimal brain strategy by determining the optimal networks and involving the external inputs. This sophisticated architecture made him more effective than traditional regression models such as AR, MA, ARIMA. In the energy management field, this algorithm is often used owing to its great abilities in time series dependencies analysis for prediction purposes. The equation of the NARX is defined as follows in equation (5). Figure (1) presents the configuration of this model.

$$y(n+1)=f[y(n),...,y(n-d_y+1);u(n),u(n-1),...,u(n-d_u+1) \qquad (5)$$

Where:

| | |
|---|---|
| $u(n)$ | Input of the model at discrete time step n |
| $y(n)$ | model output at discrete time step n |
| $d_u \geq 1$ | Input memory order |
| $d_y \geq 1$ | Output memory order |

In the standard NARX network, we have a two-layer feedforward network, with a sigmoid transfer function in the hidden layer and a linear transfer function in the output layer. This network has a specific feature by involving a tapped delay lines to store previous values of the x(t) and y(t) sequences. Note that the output of the NARX network, y(t), is fed back to the input of the network (through

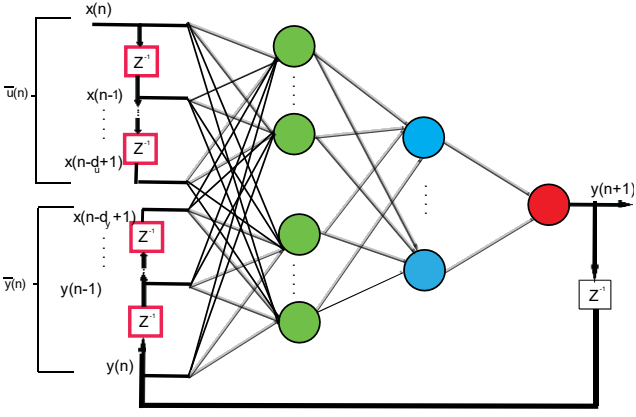

Fig. 1. NARX recurrent neural network architecture

delays). This opens a window on two different modes for training this powerful class of dynamic models:

Series-Parallel (SP) Mode: which takes the feedback delayed information from the real values given in the database used for the supervised training:

$$\hat{y}(n+1)=\hat{f}\ [\ y_{sp}(n);u(n)]=\hat{f}[y(n),...,y(n-d_{y+1});u(n),...,u(n-d_{y}+1)]\quad (5)$$

Parallel (P) mode: where the estimated outputs are set for the output's regressor:

$$\hat{y}(n+1)=\hat{f}\ [\ y_{p}(n);u(n)]=\hat{f}[\hat{y}(n),...,\hat{y}(n-d_{y+1});u(n),...,u(n-d_{y}+1)]\quad (6)$$

However, for better accuracy and effective training firstly, we use the NARX-SP feedback on open-loop then we switch to the parallel feedback in the evaluation part with a closed-loop.

Given from articles cited, the simulations prove that NARX networks give better accuracy in discovering the behavior of the time series output than conventional outputs[37]. These features are obtained from the fact that input vectors are inserted through two tapped-delay lines from the input-output signals. This is clearly mentioned from equation (1) -(3) from the parameter $d\ dx$ and $dy$ a stocking the information to reconstruct the states of the neural network. Furthermore, these delays present a jump-ahead connection in the time-unfolded network to provide the ability to the gradient descent back propagates in a shorter path and decrease the network vanishing issue in long-term prediction.

Unfortunately, when applying the NARX model in time series dependencies, the output memory will be eliminated and thus, the computational resources of these models will be significantly reduced.

As mentioned in the introduction, the particular topic of this paper is the issue of nonlinear time series prediction with the NARX network. In this type of application, the output-memory order is usually set $dy=0$, thus reducing the NARX network to the TDNN architecture presented in Eq.7:

$$y(n+1)=f[u(n),u(n-1),u(n-1),...,u(n-d_{u}+1)]\quad (7)$$

Furthermore, the vanilla recurrent neural network class suffers from the vanishing gradient problem i.e. the neural network after a specific input number stops learning and negatively affects the prediction accuracy. This problem comes when the gradient descent shrinks in long-range dependencies.

## II. LONG SHORT TERM MEMORY

LSTM cells introduced by S. Hochreiter & J. Schmidhuber[36] belong to recurrent neural networks architecture targeting the vanishing gradient problem. Since 2006 The aforementioned technique become widely used in various areas such as speech recognition, handwriting[38]-[39], weather forecasting[40]. However, the idea is quite simple, these cells by forgetting the noisy information that misleads prediction techniques and keeps only the important information to be forwarded to the hidden layers. This process is established by three gates. LSTM gates equations are given below:

$$i_{t}=\sigma(w_{i}\left[h_{t-1},x_{t}\right]+b_{i})\quad (8)$$

$$f_{t}=\sigma(w_{f}\left[h_{t-1},x_{t}\right]+b_{f})\quad (9)$$

$$o_{t}=\sigma(w_{o}\left[h_{t-1},x_{t}\right]+b_{o})\quad (10)$$

With:

| | |
|---|---|
| $i_{t}$ | Represents the Input gate |
| $f_{t}$ | Represents the forget gate |
| $o_{t}$ | Represents the output gate |
| $\sigma$ | Represents the sigmoid function |
| $w_{i}$ | Wight for the representative gate (x) neurons |
| $h_{t-1}$ | Input at current timestamp |
| $b_{x}$ | Biases for the respective gates (x) |

## III. PROPOSED ARCHITECTURE

The proposed model is a combination of tow efficient models. The ensemble model merges the properties of individual predictors to create a stronger predictor. The main feature that adds to the hybrid model from NARX network is these embedded memories that provide jump-ahead connections in the time-unfolded network. Associated with LSTM memories. These jump-ahead connections provide shorter paths for propagating gradient information, thus reducing the sensitivity of the network to long-term dependencies.

We hypothesize that additional tapped time delays from NARX with deep learning LSTM-RNN will improve the accuracy and help to prevent the overfitting from long term dependencies that can be achieved in other classes of recurrent neural network architectures by increasing the orders of embedded memory. It should be pointed out that our embedded memory simply consists of simple tapped delayed values to various neurons and not more sophisticated embedded memory structures.

The proposed framework for long-term prediction is decomposed from tow steps: NARX network receives the weather information to primarily predict the PV power. The latter is added to the original database to pass to LSTM –RNN. The output of the aforementioned model presents the final result. The reason of choosing NARX networks is due to its success on problems including the latching problem and nonlinear system identification in one side and decreasing the vanishing gradient by including LSTM memories, we explored the ability of other recurrent neural networks associated on LSTM memory to solve problems that involve long-term dependencies. The model ensemble is shown in Figure 2.

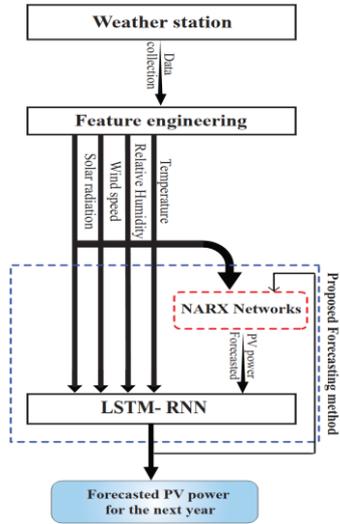

Fig. 2. NARX-LSTM recurrent neural network proposed architecture

Adding to LSTM-RNN architecture, a general class of regression model with time delay has been given prime importance in this study. it has been pointed out that a neural network could be stabilized or destabilized by certain stochastic inputs. Hence, it is significant to consider stochastic effects on the stability property of the delayed neural networks.

NARX-LSTM recurrent neural network associates two modules types: (1) NARX recurrent network that receives the sequence of external inputs as well as the recurrent output layer state. (2) LSTM cells that receive additional information with the original features in order to make a classification depending on their impact on the performance scores and then maps the internal activation function to set the outputs. Eq. 1 and 2 presents the proposed model.

$$y' = g(\sum_{i=1}^{n} u_i) \qquad (11)$$

$$Y(t) = f(y', \sum_{i=1}^{n} u_i) \qquad (12)$$

Where $y'$ is the primary output of NARX model, and $Y$ is the final PV power, $u_i$ are the weather features and $n$ is the number of inputs.The latters include the irradiance, the wind direction, the temperature and the relative humidity. While g and f are the characteristic functions of NARX and LSTM respectively. The data processing is presented in Figure 3.

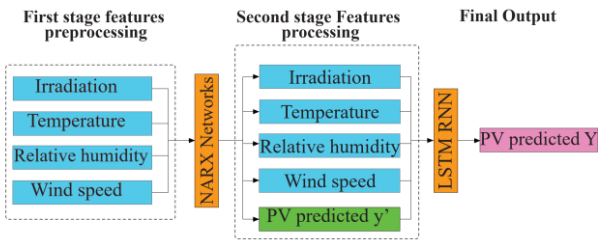

Fig. 3. Proposed algorithm

To get a better understanding of the latter model implementation, a proceeding summary is presented through a complete flow in Algorithm 1

| Algorithm1: NARX-LSTM |
|---|
| ❖ **Input:** |
|    1.  Data acquisition $X_i = \{X_i, .., X_n\} = \{IR; T; RH; WS\}$ |
| ❖ **Output:** |
|    1.  Data splitting to 80% for training, 20% for testing |
|    2.  Train NARX model |
|    3.  Predict the PV power with NARX network |
|    4.  Add the PV power predicted to the Database i.e $X_i^{new} = \{IR; T; RH; WS; Power\}$ |
|    5.  Train LSTM model with $X_i^{new}$ |
|    6.  Validate the model with 20% of testing |

IV. CASE STUDY AND SIMULATION RESULTS

A. Feature engineering

In this study, two years of historical data were analyzed to target PV power. The inputs are ambient temperature, wind speed, irradiation, and relative humidity. The database used for the training sets is from 04/01/2016 to 04/01//2018. The forecasting outputs are for a full year from 04/01/2018 to 04/01/2019 It should be pointed out that the database is cleaned from missing values and smoothed from inreal measures. The step time is for 5 minutes. The training features for tow years are plotted in Figure 4.

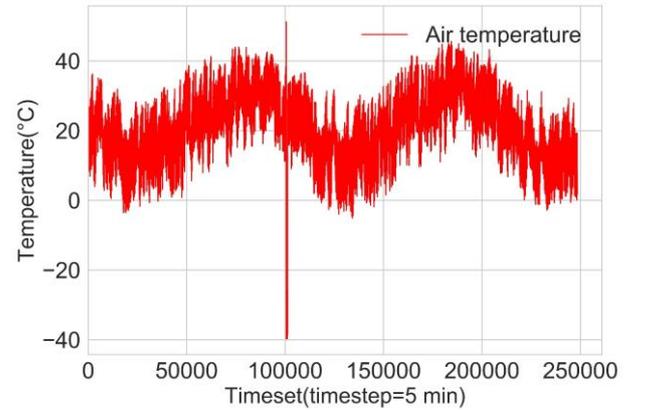

(a)

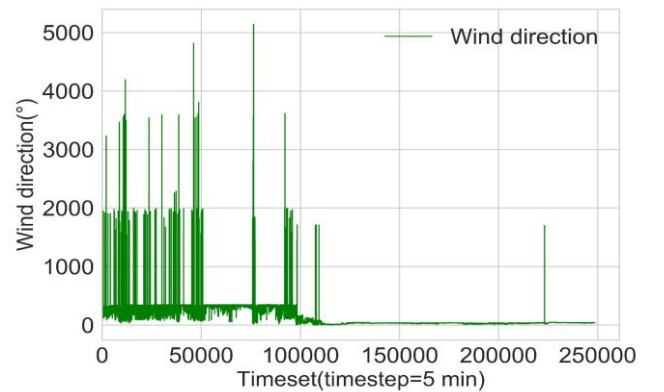

(b)

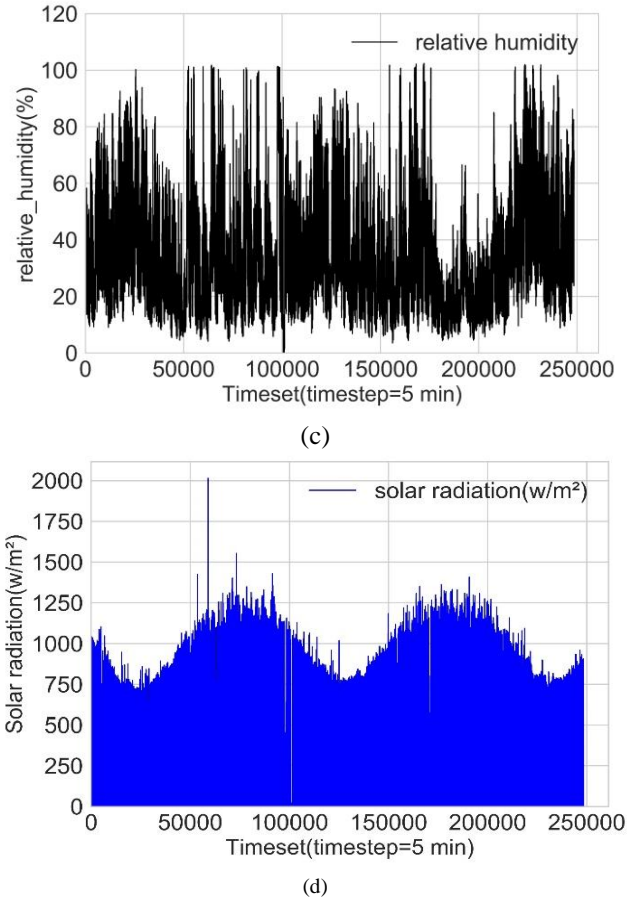

(c)

(d)

Fig. 4. Variation of weather parameters in one year: (a)Air temperature($^\circ C$) (b)Wind direction ($^\circ C$ )(c) Relative humidity(%) (d)Solar radiation(W/m²)

The inputs parameters have a direct relation to the predicted output. However, this relationship is not equally repartitioned. Various models have been proposed to measure the diversity of feature importance. The methodology is to permit one parameter and estimate the increase of forecasting error in each case. In this study, the domain knowledge is investigated using Elastic Net via ranking the features according to their relative coefficient magnitude. This method combines Lasso and Ridge models. Figure 4 presents the relative importance of each input parameter basing on the relative coefficient magnitude.

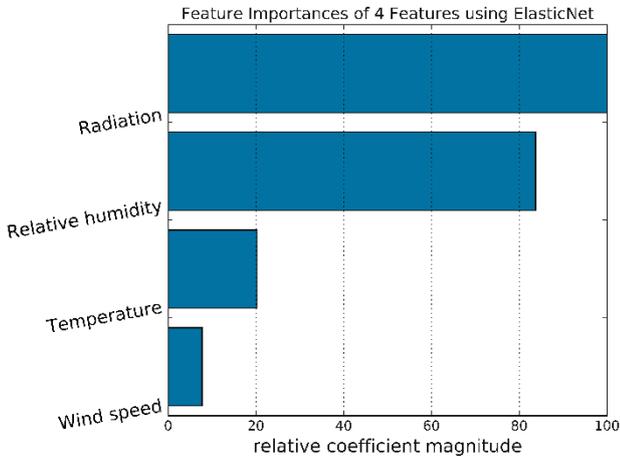

Fig 5. Relative importance for feature inputs with ELASTIC NET.

According to Elastic Net tool, the irradiation has a major part in the prediction accuracy followed by the relative humidity.

The evaluation of the proposed model is done through three parts: Cross-Validation, scores metrics comparison, and real/forecast plots. Splitting the data into prediction and testing is essential to assess the fitness of the model. The K fold Cross Validation split the training data into 10 folds. 10 used for the training and the latest part is used for testing. By using this approach, each fold is involved in the assessment at some point. Due to the large dataset used in the training, this step is a time consuming, However, it gives high reliability for the model testing with fixed scores metrics. The score evaluation functions used in this model are the rooted mean square error function, the mean absolute error as well as the mean percentage error function. The equations of the said error functions are written as following in 13, 14. Table 1 present the result of 10-folds Cross-Validation.

$$MAE = \frac{1}{n} \sum_{i=1}^{n} \left( \widehat{y}_i - y_i \right) \qquad (13)$$

$$RMSE = \sqrt{\frac{1}{n} \left( \sum_{i=1}^{n} \left( \widehat{y}_i - y_i \right)^2 \right)} \qquad (14)$$

$$MAPE = \frac{100\%}{n} \sum_{i=1}^{n} \left| \frac{\widehat{y}_i - y_i}{y_i} \right| \qquad (15)$$

Table. 1. Predicted vs. Actual PV power in one year

|  | RMSE | MAE |
|---|---|---|
| CV with 10 folds | 11.2 | 5.67 |

From Table 1, It can be mentioned that RMSE=11.2 which refers to high accuracy. A MAE= 5.67 confirms the high accuracy of the proposed model. This latter is computed after an automatic tuning using Randomized Search module. The input parameters are normalized in order to increase the convergence speed and the performance of the model. Thus, the minimum-maximum scalar with a magnitude range of [0, 1] is applied to features values.

As a sequence of inputs, the features are introduced in the recurrent neural network. The batch size, the number of epochs and the learning rate are fixed through hyperparameter optimization tools. The inputs are the weather parameters such as irradiance, temperature, wind speed, and relative humidity. We used three LSTM layers. And the activation function used is sigmoid function and the last layer uses a smooth approximation to the arg maximum function Softmax as an activation function. The proposed model is a NARX - LSTM recurrent neural network model. The ability of LSTM memory cells to save important information as preventing the vanilla networks model makes the model gratefully fitting the PV power outputs. Moreover, the NARX networks made the hybrid model efficient in time series PV power forecasting for longer range prediction.

The weather data of 2016- 2017-2018 are used in the training and the testing sets. In this experiment, tow years are used for the training and one year is used for testing purposes. The rich database for a small-time step of 5 minutes made the analysis more exhaustive. The large database is fed in the system due to the high variability of the input parameters from one year to another.

With the aim of assessing the performance of the new model, the NARX-LSTM is simulated for one year, then decreasing the time steps to one month and one day to show clearly the behavior of the proposed model in Figures 6-7 and 8.

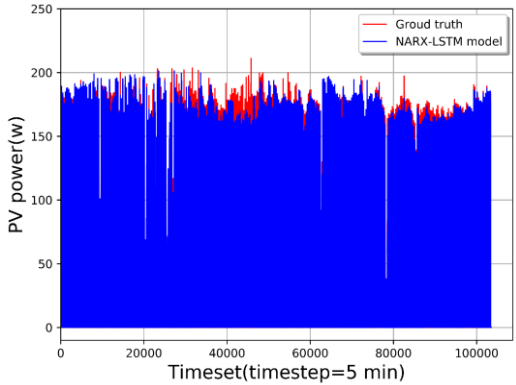

Fig. 6. Predicted vs. Actual PV power in one year

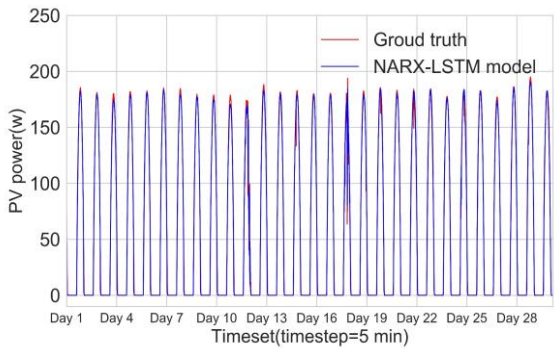

Fig. 7. Predicted vs. Actual PV power in one month.

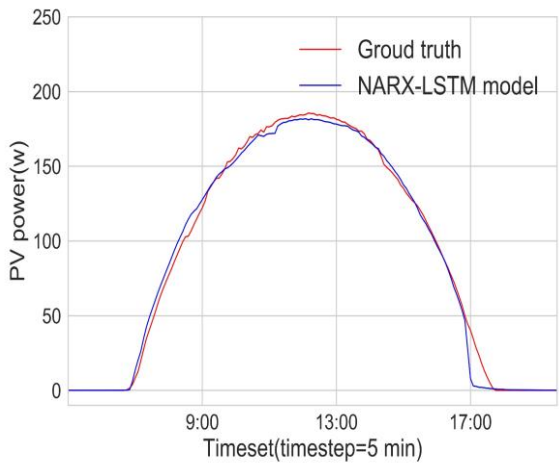

Fig. 8. Predicted vs. Actual PV power in one month.

Figure 6,7,8 present the shapes of the PV power from the actual/predicted values. The estimated power follows the real one in an impressive way. From the mixed points in blue which envisione the sole of NARX-LSTM with the red points of the real values, the error rate is low. It can be concluded that the proposed method is notably efficient with long term dependencies. This result comes from figures 7 and 8 that illustrate the effectiveness of NARX

networks in decreasing the error rate from the concatenated ensemble model. The loss functions and the mean squared error in the training phase are shown in Figures 9 and 10.

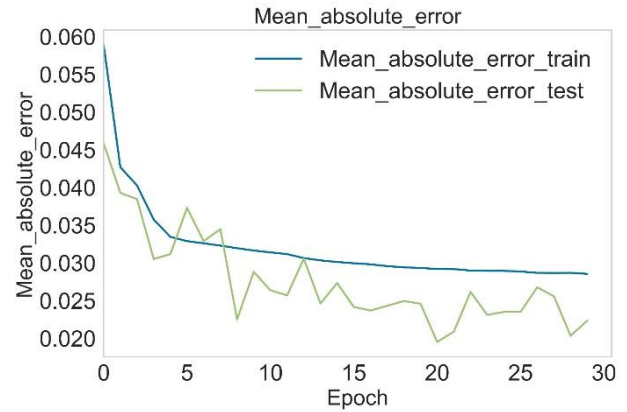

Fig 9. MAE function visualization during the training step

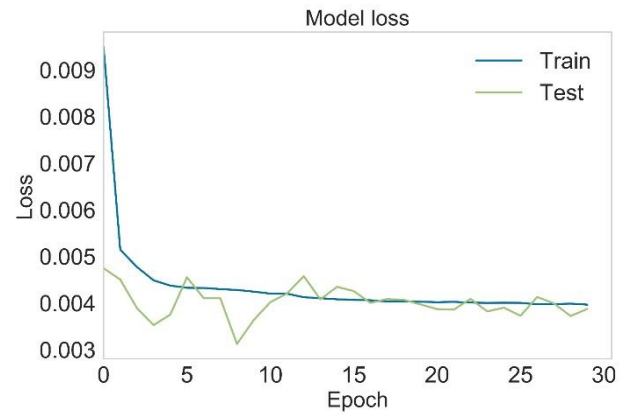

Fig 10. Loss function visualization during the training step

The error values from the loss function and the MAE are decreasing significantly from 0.06 from the first step to 0.02 in the ending step for the loss function and for 0.009 to 0.003 in for MAE score. It should be mentioned that these values are scaled between [0,1]. The scores metrics including RMSE, MAE, and MAPE are envisioned from Table 2. The lower they are, the higher the accuracy the model performing.

Table 1. Performance metrics from 04/01/2018 to 04/01/2019.

| Score function | RMSE | MAE | MAPE |
|---|---|---|---|
| Value | 10.51 | 4.72 | 0.269% |

The model is simulated for 46 minutes with 30 epochs. This is considered time-consuming However the results are worthy. In terms of accuracy, the proposed model provides a good precision the RMSE calculated is 10.51 while the MAE is 4.72. These results present a performance in handling time for series forecasting.

Furthermore, For a better assessment, a fair comparison is made with the benchmarked models. The forecasting techniques include Extra Trees regressor, LSTM-RNN and K Nearest Neighbors (KNN) [41],[42]. We choose the original LSTM-RNN to see the contribution made in terms of accuracy for the proposed model. Moreover, Extra tree regressor named also Extremely randomized trees presents a branch from Boostrap aggregation(Bagging). The latter uses an ensemble model to build from weak learners a robust predictor. The said method yielded state-of-the-art results with high variant features. The last model interpreted is KNN. This aforementioned method uses $k$ samples to the unknown labels and calculates their average. KNN is frequently used in statistics and revolutionary computing for the high efficiency it provides.

The aforementioned models are tuned and trained using the same inputs parameters with a hyperparameter Randomized Search for a fair comparison. The scores errors interpreted for each model are the RMSE and MAE. Figure 11 presents the summary of all the models vs the real PV power. Consequently the error values are collected in Table 3.

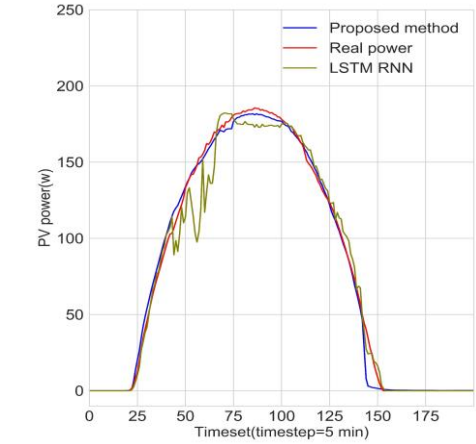

(a)

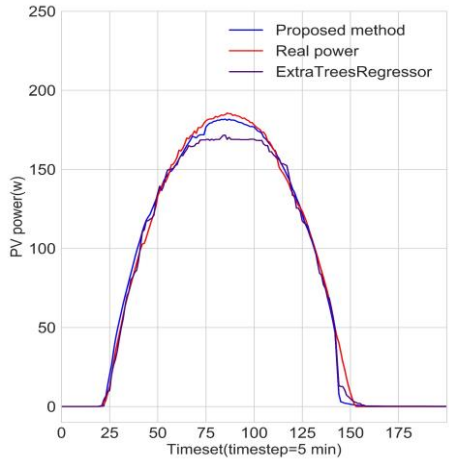

(b)

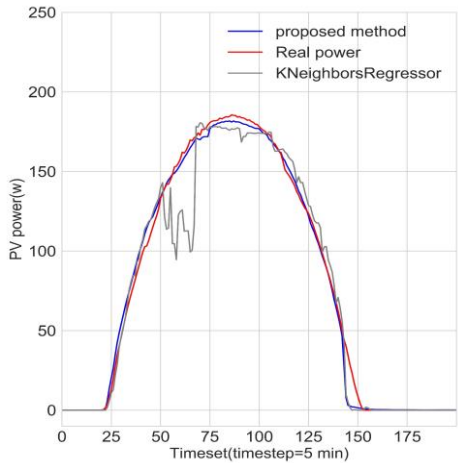

(c)

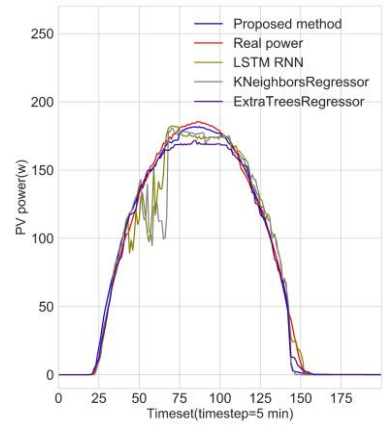

Fig.11. NARX-LSTM and real PV power plot simulation compared with (a) LSTM-RNN (b) Extra Tree Regressor (c) KNN

Table 2. Score performance comparison after experimental results using Randomized Search for hyper parameter tuning

| Model | RMSE(W) | MAE(W) |
|---|---|---|
| Extra Trees | 7.98 | 4.49 |
| KNN | 9.73 | 5.11 |
| LSTM RNN | 7.41 | 4.26 |
| NARX-LSTM | 6.27 | 3.81 |

From Figure 11, each method is indicated by its specific color. The cyan, grey, violet indicate LSTM, KNN, and extra tree regressor respectively while the red and the blue present the real power and NARX-LSTM outputs. It can be noticed that the proposed technique is outperforming the cited models from the mixed points between the real values and the forecasted results. This result is confirmed by registering the lower RMSE and MAE for the proposed method with 6.27 watts of RMSE and 3.81 watts of MAE. An improvement of 15.38% is made. The PV power estimated from the aforementioned techniques as well as the proposed technique presents a high correlation between the actual/observed values. The model smoothly follows the real data. This amplifies that the said approach made it the most accurate compared to the aforementioned models. This proves that the gradient descent learning is not shrunk in NARX-LSTM networks. Moreover, the said model can follow even the sudden spikes which made obviously the strength of our model is its ability to capture the spikes generated from the sun more than any predictor. However, the computing time is creating a serious issue if we increase the number of epochs to get these values. We can say that the training step is the most relevant drawback for this model. The simulation took 2 hours 14 minutes for 50 epochs in a LENOVO Ideapad 720S-15IKB (i7 with 8 Cpu). Even with the use of GPU parallel processing provided by NVIDIA, the training is a time consuming compared with the benchmarked methods. This may lead to some lagging especially if the model is used in online training. Nonetheless, the proposed provides a reliable tool to handle the vanishing gradient for long term forecasting.

# V.Conclusion

This paper presents NARX-LSTM model in medium-long term photovoltaic power forecasting. The said model is high performing in time series prediction with an RMSE of 6.27. Furthermore, this model is the relevance of nonlinear autoregressive with exogenous inputs and LSTM cells. Memory cells protect the gradient from vanishing issue while NARX networks ensure the long-range prediction through its architecture. The inputs entered are the temperature, solar irradiance, wind speed and relative humidity which create a great combination with the necessary variance to ensure an accurate prediction from the stochastic climate change. The said model is tested on a rich database of an Australian plant and compared to a various group of models.

Regarding the various simulations, the said model presents a good performance in comparison with the common methods used in regression aims. From table 3, We found that this NARX-LSTM network is the most accurate in terms of RMSE.

The said model can be used in long term forecasting for unit commitment and budget planning in long-range with a high certitude. However, the extensive computational work made the simulation time slow comparing to the techniques used in the prediction so, there is a need for a high performing laptop to accelerate the training time especially if there is a need to do online supervised learning. Another suggestion propose an extensive feature engineering to optimize the database and then reduce the samples entered to LSTM gates. The outcome of this study comes from the combination of regressive and machine learning algorithms to the PV power forecasting. This model has the ability to capt the behavior of the temporal weather changes and transform it into PV power energy easily. In this respect, the combination of regressive and machine learning technique in time series forecasting presented by the proposed technique in this study opens the door for an accurate long term forecasting with promising results in the future.

## Acknowledgments

The authors would highly acknowledge the financial support of the Qatar National Research Fund (a member of Qatar Foundation). Also special thanks to Pr. Haithem Abu-Rub from Smart Grid Center Laboratory SGC who made this study possible.